\documentstyle[pre,aps,prb,a4,epsfig]{revtex}

\begin{document}
\input epsf
\draft
\title{A molecular dynamics simulation of
water \\ confined in a cylindrical $SiO_2$ pore.  }

\author{M. Rovere\footnote[1]{Author to whom correspondence 
should be addressed}, M.A. Ricci, D. Vellati and F. Bruni }
\address{Dipartimento di Fisica, 
Universit\`a di Roma Tre\\
Via della Vasca Navale 84, 00146 Roma, Italy. \\
INFM, Unit\`a di Ricerca Roma Tre}

\author{ }
\address{ }
\maketitle

\begin{abstract}
A molecular dynamics simulation of water confined in a silica pore is 
performed in order to compare it with recent experimental results
on water confined in porous Vycor glass at room temperature.
A cylindrical pore of $40 \AA$ is created inside a vitreous $SiO_2$ 
cell, obtained by computer simulation. The resulting cavity  
offers to water a rough hydrophilic surface and its geometry and size are 
similar to those of a typical pore in porous Vycor glass.  
The site-site distribution functions of water inside the pore
are evaluated and compared
with bulk water results. We find that the modifications of the site-site
distribution functions, induced by confinement, are in qualitative agreement 
with the recent neutron diffraction experiment, confirming that the
disturbance to the microscopic structure of water mainly
concerns orientational arrangement of neighbouring molecules.
A layer analysis of MD results indicates that, while the geometrical
constraint gives an almost constant density profile up to the layers
closest to the interface, with an uniform average number of
hydrogen bonds (HB), the hydrophilic interaction produces the wetting
of the pore surface at the expenses of the adjacent water layers.
Moreover the orientational disorder togheter with a reduction of
the average number of HB persists in the layers close to the interface,
while water molecules cluster in the middle of the pore at a
density and with a coordination similar to bulk water.
\end{abstract}



\newpage

\section{ Introduction}
Recent literature reports on many studies about the structural and dynamical
properties of water confined in different environments.~\cite{chen} The
interest on this subject arises mainly from evidence of the relevance of the
water-substrate interaction in determining either stability and enzymatic
activity of proteins or swelling of clay minerals. In this respect it is
also important to elucidate to what extent water itself is significantly
perturbed from its bulk behavior, when confined within a porous material,
and in particular how far from the interface does the perturbation propagate
through the liquid and whether it depends on the nature of the interaction
(hydrophilic or hydrophobic).

All the experimental studies of the dynamics of water molecules confined in
different substrates agree in suggesting a slowing down of the translational
single molecule motion~\cite{chen,10I,11I} compared to the bulk liquid phase,
while the same kind of agreement is not found in the interpretation of
structural data.~\cite{2I,3I,7I,14I,I,II} Most neutron diffraction studies
are indeed interpreted in terms of more extensive hydrogen bonding than in
bulk water, at least in the water layers closer to the interface and 
below room temperature~\cite
{2I,3I,7I,14I}, contrarily to what found in a recent study performed by some
of us.~\cite{I,II} As a matter of fact this study differs from previous ones
in two relevant aspects: first, only in this case hydrogen/deuterium
substitution in the neutron diffraction experiment was exploited, thus
allowing the estimate of three site-site radial distribution functions;
second, full account of the excluded volume effects on the radial
distribution functions was attempted for the first time. Although the
limitations of the experimental technique do not allow to draw 
definitive and quantitative
conclusions, nevertheless it was apparent from this study that water
confined in porous Vycor glass is still hydrogen bonded, but
the bond network is strongly distorted even at room temperature. 
The above mentioned experimental limitations
are due to the presence of contributions to the measured cross section due
to the cross correlations between water atoms and Vycor atoms, which cannot
be isolated from the water-water correlations in the neutron
diffraction experiment. As a
consequence in the experiment only three composite
site-site distribution functions are accessible through the isotopic
substitution, namely $g_{HH}(r)$, $g_{XH}(r)$, and $g_{XX}(r)$, where the
subscript $H$ stays for all hydrogen atoms in the sample (either water
hydrogens or protons bonded to the pore surface)
and $X$ labels all non substituted atoms 
(i.e. water oxygens or Vycor atoms).~\cite{I,II} 
Information about the relative arrangement of water atoms may
then be masked by the cross correlation terms: to extract this information
from the experiment the hypothesis that the microscopic structure of
confined water does not change upon changing the hydration state was
proposed.~\cite{II} 
Molecular Dynamics (MD) simulations on the contrary do not
suffer from such limitations, as individual atomic species may be labelled
and the corresponding distribution functions calculated, thus providing an
useful test of the entire data analysis procedure.

To our knowledge, only a few MD studies on the modifications 
of water structure and/or
dynamics in the presence of a substrate interaction have been so far
performed~\cite{rossky1,rossky2,rossky3,zhang,spohr}, 
for both hydrophobic or hydrophilic
interactions. All these studies indicate that water perturbations are both
short ranged and relatively mild in magnitude. However the investigated
geometries may hardly be comparable with the extended 
and fractal-like network of thin
cylindrical pores offered by Vycor glass. Since the nature of the
substrate-water interaction along with the geometrical characteristics of the
confinement may influence the structure of interstitial water\cite{soper},
we present here a MD simulation of TIP4P water, confined in a
cylindrical pore built inside a SiO$_2$ glass.

\section{ Molecular dynamics of water in a silica cavity }

To model the cylindrical cavity of a porous glass, like Corning Vycor
glass~\cite{vycor}, we follow the
method proposed by Brodka and Zerda~\cite{brodka} in the simulation
of liquid acetone in silica pores. A vitreous silica (SiO$_2$)
of 192 atoms is obtained by performing a MD simulation
with the empirical potential model of Feuston and Garofolini.~\cite{feuston} 
Starting from a cristobalite crystalline structure~\cite{crist}, the
system is melted at $6000 K$ and then quenched to room temperature
with the method described in 
Ref.~\onlinecite{brodkaold}~and~\onlinecite{brodka}. Since
we need a cylindrical cavity of $40 \AA$ of diameter, which corresponds to
the average size of the Vycor glass pores, the small glass cell is repeated
five times along the three directions in order to get a cubic cell
of approximately $19000$ atoms  with a box lenght of 
$L=70 \AA$. We create a cavity inside the glass by
removing the atoms lying within a distance $R=20 \AA$ from the $z$-axis.
Then we remove from the cavity surface also the silicon atoms bonded 
to less than four oxygens. This procedure
leaves on the pore surface a number of {\em non bridging} oxygens (nbO),
i.e. oxygen atoms bonded to only one silicon, and is a reasonable
replica of the industrial preparation of the sample.~\cite{vycor}
Surface oxygens bonded to two silicon atoms will be in the
following referred to as {\em bridging} oxygens (bO). This nomenclature 
is in agreement with previous literature.~\cite{brodka}
Since in the experiment the nbO were saturated with protons prior to
hydration, we attach extra hydrogen atoms to the nbO found on our simulation
cell surface. The surface hydrogens 
are placed at the equilibrium distance ($0.95 \AA$) from a nbO
with a Si-O-H angle of $116^o$. 
We notice that the procedure described above generates a cavity
with a rough surface, with geometry and size similar to the average ones
in the real porous Vycor glass. The actual 
volume of the cavity, $V_p$, generated
inside our simulation box is indeed unknown and only roughly approximated to
a lower value by the volume of a cylinder of radius $R=20 \AA$.

In the simulation we use the model TIP4P~\cite{tip4p} for water. 
Water molecules interact
with the substrate atoms with a potential 
modeled according to Ref.~\onlinecite{brodka},
where different Lennard-Jones (LJ) parameters
and fractional charges are assumed for bO and nbO. 
Silicon and surface hydrogen atoms interact only 
via coulombic forces with the charged water sites. LJ parameters and
fractional charges for the $SiO_2$ sites and TIP4P water are given in Table~1.
The cross LJ parameters are calculated
with the usual Lorentz-Berthelot rules. During the simulation
the substrate atoms and the surface hydrogens do not move.  
Periodic boundary conditions are
applied only along the axis of the cylinder ($z$-direction).
All the pair interactions are truncated at a cutoff radius of $r_c=9 \AA$ and 
reaction field corrections are applied.~\cite{allen,neumann}
 
The number of particles to use is determined by the density of
the confined water at full hydration in the neutron diffraction 
experiment~\cite{ross}, i.e. $n=0.0297A^{-3}$ corresponding to
$N=2661$ molecules. 
The simulation starts with the centers of the
molecules placed in a fcc lattice, where the occupied nodes are randomly
chosen. The size of the lattice is such that it is contained in the
cylindrical pore. The origin of the coordinates is assumed in the 
middle of the cylinder axis. The initial velocities 
are chosen randomly with a distribution consistent with the required
temperature. The equation of motions are solved with the 
popular quaternion-leapfrog algorithm due to Fincham.~\cite{allen}
Confined water is melted at $500 K$ and then equilibrated
at $300 K$.
Initially a very small time step of $10^{-4}ps$  is used then it is
slowly increased. We find that for time steps larger than $5 \times 10^{-3}ps$ 
we need a too frequent rescaling of the temperature, 
so the final runs are done with a time step of $10^{-3}ps$.
The calculation of the forces during the simulation is performed with  
a neighbour list built by using a cell method~\cite{furio}, 
thus allowing the simulation to be run on an Alpha station.

\section{ Site-site distribution functions}

The calculation of the site-site radial distribution functions of confined
water is motivated by the possibility of comparing with the experimental
results recently obtained with the use of the isotopic substitution 
technique in neutron diffraction experiments.~\cite{I,II} 
On the other hand the site-site distribution
functions can be easily calculated from computer simulation, when periodic
boundary conditions are applied in all three directions, but in our case,
where the system is confined in a cylindrical box one must carefully
take care of the excluded volume effects, as discussed 
recently by Ref.~\onlinecite{soper}~and~\onlinecite{II}. As a matter of fact
the radial distribution functions obtained either experimentally or
by simulation for a confined liquid cannot readily compare with those
of the corresponding bulk, due to the existence of regions, where
the liquid is not allowed. In this case indeed the radial distribution
function of a system of non interacting particles, the so called {\sl
uniform fluid}, is not equal to unity at all $r$ values, but approaches its
asymptotic value with a profile which depends
on the size and shape of the confining volume. As a consequence the
site-site distribution functions of the interacting liquid, which are
defined with respect to that of the uniform fluid, do not lay on a flat
profile and the amplitude of their oscillations may also show
a trivial $r$-dependence.~\cite{soper,branka}

Our MD results could directly compare with the experimental data of 
Ref.~\onlinecite{II} only if the simulation box was topologically identical to 
the real Vycor glass sample and if the site-site distribution functions
of TIP4P water were correctly reproducing those of real bulk water. 
Since both conditions are only approximately matched, we will perform
the corrections for the excluded volume effects on our MD calculated
site-site distribution functions, to compare the corrected ones with
those of TIP4P bulk water. The aim of this comparison will be to test 
whether we observe the same qualitative modifications with respect 
to bulk water observed in ref.~\onlinecite{II}.

From the configurations obtained in the molecular
dynamics simulation one can calculate for each site of type $\alpha$,
placed in the origin of the reference frame, the
number of sites of type $\beta$ lying inside a spherical
shell $\delta v(r)$ of thickness $\delta r$ at distance
$r$ from the origin and obtain the average number of site pairs 
$\overline{n}^{(2)}_{\alpha\beta}(r)$. Following the
usual recipes~\cite{allen}, regardless of the confinement
inside an almost cylindrical volume,
the computer simulated distribution function is then calculated as
\begin{equation}
g^{MD}_{\alpha\beta}(r) = \frac{\overline{n}^{(2)}_{\alpha\beta}(r)}{
{\frac{N_\beta}{V_p}}\delta v(r)} \quad \label{defgmd}
\end{equation}

where $N_\beta$ is the number of $\beta$ sites.
The results obtained for the $g^{MD}_{\alpha\beta}(r)$ 
functions are shown in Figs.~1a~-~1c.
They are compared with the same functions obtained in a simulation, where
water is purely confined in the cylindrical pore without any interaction
with the substrate atoms. The large difference between the intensity of the 
first peaks of the two oxygen-oxygen $g^{MD}_{OO}(r)$  
functions indicates that switching on the interaction
with the substrate atoms is equivalent to enhance the confinement effects. 
The same trend is observed for the other functions and particularly 
in the oxygen-hydrogen $g^{MD}_{OH}(r)$ one, 
where the so called hydrogen bond peak at $r \approx 1.85 \AA$ is strongly 
enhanced. However,
as we will see below, this is not an evidence of an increase
of the number of hydrogen bonds in the confined system, but a trivial
effect of the geometrical confinement.~\cite{branka}

From the Figs.~1a~-~1c it is clear that in the limit of large $r$ all three
site-site correlation functions of confined water go below 
the exact limit of unity. This is a finite size effect
due to the absence of periodic
boundary conditions along the $x$ and $y$ directions.

As already stated the site-site  correlation
functions ~(\ref{defgmd}) calculated in the computer simulation
represent the fluctuations of the density with 
respect to a
{\em uniform} density profile, which is not a constant as in
non confined systems. Thus to obtain 
the corrected site-site 
distribution functions of water inside
the pore $g_{\alpha w\beta w}(r)$ 
to be compared with those of the corresponding bulk liquid, we
calculate, following Ref.~\onlinecite{soper}~and~\onlinecite{II}:
\begin{equation}
g_{\alpha w\beta w}(r) =\frac{g^{MD}_{\alpha\beta}(r)}{g^{ww}_u(r)}
  \quad \label{gwwdef}
\end{equation} 
where $g^{ww}_u(r)$ is the {\em pair correlation function} of the
confined uniform fluid.

At variance with the experiment, in the present computer
simulation, water is confined in a single pore, thus
neglecting the correlation between water molecules residing in
different pores, nevertheless the molecular
dynamics results can be corrected using the same basic results of
Ref.~\onlinecite{soper}~and~\onlinecite{II}. 
In particular the function $g^{ww}_u(r)$ can be obtained from 
the Fourier transform of the form factor of a cylinder.~\cite{kratky}
Moreover to evaluate eq.~(\ref{gwwdef}) one has to take into account that
the function $g^{MD}_{\alpha\beta}(r)$
is not known behind $r_c$ and its asymptotic behaviour
is not reliable, a problem already considered in the experiment
where the $q \rightarrow 0$ limit of the structure factors are
not available.~\cite{II} 

Before commenting the MD results after correction for excluded volume effects,
and comparing these with bulk water results, it is useful to remind 
the qualitative differences between bulk and confined water evidenced
in Ref.~\onlinecite{II}. 
The water-water distribution function for the oxygen atoms
of confined water $g_{O wO w}(r)$ has the first peak at $2.8 \AA$ as in
the bulk, although sharper and less intense, while the second peak is shifted
towards $3.4 \AA$. Only minor differences are apparent in the distribution 
function of water oxygens and water+Vycor hydrogens, $g_{O wH {w+s}}(r)$, with
respect to the $g_{OH}(r)$ of bulk water. In particular the H-bond peak
has almost the same intensity in both liquids,
although in the confined case it is sharper, and some excess intensity 
is visible between the two main peaks. 
The hydrogen-hydrogen distribution function of water sites, 
$g_{H wH w}(r)$, shows on the contrary a dramatic increase of the first peak 
at $2.3 \AA$, while the second one is almost completely
washed out.
All these results suggest that confined water is still H-bonded,
nevertheless the orientational 
correlations between neighbouring water molecules are strongly influenced
by the confinement, resulting in a highly distorted network of bonds. 
As far as the water-Vycor cross distribution functions are concerned,
only the distribution functions of water oxygens and Vycor atoms 
(regardless of any distinction between oxygen and silicon), $g_{O wX s}(r)$,
and that of Vycor atoms and water hydrogens, $g_{X sH w}(r)$, 
were extracted from the experiment. The first one shows an intense peak
at $2.8 \AA$, followed by a second peak at $4.5 \AA$; the second one shows
modulations, with very low intensity at $2 \AA$ and $3.4 \AA$.

In Fig.~2a~-~2c we show the results obtained  with a procedure
similar to the one adopted in Ref.~\onlinecite{II}, in order to perform
the corrections for the excluded volume effects.  
Due to the approximations involved in the derivation, our
functions are not reliable below the minimum approach distances, evidenced
in Fig.~2a~-~2c by the arrows. 
The site-site correlation
functions of the bulk TIP4P water at the density $n=0.0297 A^{-3}$ are
also shown in the same figures: although this density does not correspond
to a physical state for bulk water at ambient conditions, nevertheless 
we prefer this state as a reference, to avoid confusion of density effects with
those due to confinement.

The oxygen-oxygen site-site distribution function (Fig.~2a) 
shows a first peak, which
is lower in amplitude and sharper than in bulk water, in agreement 
with the experiment. Moreover
there is an evident
increase of intensity in the region between $3-4 \AA$: an indication
of a distortion of the hydrogen bond network found also
in the experiment.~\cite{II}
As far as the $g_{O wH w}(r)$ function is concerned, we notice a 
dramatic change of the amplitude of the second peak and of the first
minimum, in comparison with bulk water, while only minor
modifications of the H-bond peak seem to occur. This results
suggest that while the average number of H-bonds, as measured by
the integral under the first peak, does not sensibly change, the
orientational order of neighbouring water molecules
is strongly distorted in the confined geometry. The comparison
with the experiment is not straightforward in this case, since neutrons
cannot distinguish between water protons and protons bonded to nbO.
Nevertheless we notice that also the experimental data do not evidence
enhancements of the amplitude of the H-bond peak and that the
disagreement found with our simulation as far as the amplitude of the
second peak is concerned may, at least partly, be due to the
protons on the cavity surface, contributing with a broad peak
between $2$ and $4 \AA$ (see Fig.~3a).
Moreover in Fig.~2b we notice that
the peak around $7 \AA$, i.e.
where the surface hydrogen contribution is already very low
according to Fig.~3a, is 
in agreement with the experimental results.
 
The $g_{H wH w}(r)$ function (Fig.~2c) confirms the strong distortion of
the orientational order, since the amplitudes of both first
and second peaks decrease and the first minimum shifts towards lower 
$r$ values. On the other hand this is the distribution function with
the worst agreement with the experiment.
Such disagreement may be ascribed to the potential model or to the
approximation made in Ref.~\onlinecite{II} to extract the water-water
contribution from the measured structure factors (i.e. weak dependence
on the hydration state). 

The site-site distribution functions of the substrate atoms with respect 
to the water atoms can be obtained along the same procedure described for the 
$g_{\alpha w \beta w}(r)$ using eq.~(\ref{gwwdef}), with the appropriate 
substrate-water uniform fluid function $g^{sw}_u(r)$. Since the substrate
atoms are confined in a region between the cylinder surface and the
surface of the total simulation cell of lenght $L$, we assume that
$g^{sw}_u(r)$ is the Fourier transform of the difference between
the form factor of the cylinder of diameter $L$ and the form factor
of the internal cylinder with the diameter $2R=40 \AA$. In Fig.~3b we
show the results for $g_{O s O w}$ and $g_{O s H w}$.
The behaviour of this functions agrees with the experiment, particularly
the positions of the first peak in both functions are located
close to the experimental results.

In conclusion our simulation agrees with the experiment at a qualitative level:
both techniques suggest indeed that the H-bond network is strongly
distorted in confined water, although this does not necessarily imply strong
enhancement or depression of the H-bond peak.

\section {Layer analysis of the microscopic structure}

In order to gain a better insight into the microscopic structure of the
confined water, we perform a careful analysis of the configurations obtained
in molecular dynamics. We divide the cylindrical pore in ten concentric radial
shells, the $n^{th}$ shell being defined by
\begin{equation}
(n-1) \Delta R < \sqrt{(x^2+y^2)} < n \Delta R \quad \quad \quad n=1,\ldots,10
\end{equation}
where $\Delta R=R/10$.
First of all we look at the average density
in each shell. 

From Fig.~4 we see that the density profile is not constant
along the $r=\sqrt{(x^2+y^2)}$ coordinate, and that it takes a value
similar to that of bulk water at ambient conditions in the first layer, i.e.
the one closer to the center of the cylinder.
The density slowly decreases going 
towards the cylinder surface and abruptely increases 
at layer n.~$9$, close to the substrate surface. The last layer (n.~$10$)
is in the volume excluded by the LJ repulsion.
It is interesting to compare this result with the calculation performed with
the substrate-water interaction switched off. In this latter case the
density is essentially constant up to layer n.~$8$ and decreases in the
last two layers. The effect of the interaction is thus to attract few water
molecules, mainly from layers $7$ and $8$, to the surface, as shown by
the density increase at layer n.$9$. The disturbance, brought into the
density profile by the presence of an interacting surface, seems to
extend up to no more than $8 \AA$ from the surface itself. 
Moreover we notice that the potential model used here,
gives a modest hydrophilic effect, in comparison with 
Ref.~\onlinecite{rossky2};
this may also depend on the geometrical simmetry of the interface,
as already apparent from Ref.~\onlinecite{zhang}.
At all layers, except the $10$-th, the density is higher than its
average value of $n=0.0297A^{-3}$, due to the size of the LJ diameters.
This qualitatively explains why experimentally one finds for confined
water a lower average density than for bulk water.~\cite{ross}
On the other hand this finding needs a deeper investigation, because
it may be dependent on both the potential model and the simulation
technique adopted. In particular one should investigate whether in
a NVE simulation the system tries to build up a kind of bulk phase
at contact with the confined phase, due to the simultaneous constraints
of a constant number of particles and a constant volume.
Different computer simulation
methods, where one allows the system to change density during the
simulation~\cite{allen}, could help
to clarify this point.

In Fig.~5 we report the average
number of HB in each shell (solid line),
as calculated according to a geometrical    
definition.~\cite{brodholt} We consider two water molecules as being 
hydrogen bonded if their O-O separation is less than $3.0 \AA$, their closest
O-H separation is less than $2.3 \AA$ and the $H-O \cdots O$ angle $\gamma$ is
less than $30^o$. As seen from Fig.~5 the number of HB 
almost monotonically
decreases approaching the interface; in particular in
the layer n.~$9$, where the density is highest, the number of HB
goes below its density-weighted average value (dotted line in the figure).
When the substrate-water interaction is turned off the number of HB is
essentially constant up to the $9$-th layer and decreases,
as also the density does, in the last two layers.
The comparison of the two profiles in Fig.~5 reveals that the pure
geometrical confinement is responsible for the reduction
of the number of HB at the interface; switching on the interaction does not
sensibly modify this number, although the density at layer n.$9$ increases.
Conversely, in the layers closer to the center of the pore, where the
density reaches that of ambient water, even the number of HB 
approaches the values typical of bulk water at ambient conditions.
The reduction of the number of HB in the intermediate layers seems also
to be a density effect.  
Figs.~6a and 6b report the distributions of  HB  and
the distribution of $\cos(\gamma)$ 
respectively. The histograms of Fig.~6a broaden
and shift towards zero, approaching the pore surface,
and also the distribution of $\cos(\gamma)$ markedly broadens
for layer $9$, compared with layers $1$ and $5$. 
The pattern of HB in the $9$-th layer is severely disordered due to the 
orientational constraints, brought by the presence of a
confinement. 
We notice that the results found for the internal layers cannot be 
distinguished from those for the bulk and
the disturbance brought by the confinement in an interacting
substrate extends up to $8 \AA$ from the interface.

We have also calculated the number of HB between 
the water molecules, located
in layers $9$ and $10$, and the hydrogens and oxygen atoms belonging to 
the cavity surface. This analysis, although biased by the low statistics,
indicates that the number of such bonds is very low in
the present simulation. 

\section{Conclusions}

The results obtained by MD simulation of TIP4P water confined in a cylindrical 
$SiO_2$ pore have been compared with those of the simulation of the
same liquid after switching off the interaction with the substrate atoms and
with those of the corresponding bulk liquid.

The first comparison indicates that the absence of periodic boundary 
conditions and the cylindrical simmetry of the confining volume
are by themselves able to induce a significant distortion of the
orientational order of water molecules, although leaving the density
profile almost flat behind the shell closest to the interface.
Indeed switching on the interaction with the substrate produces only
an enhancement of the modulations of the distribution functions at
short distances, without emergence of any new feature. Moreover when the 
interaction with the substrate is switched on, the competition between
water-water and water-substrate interaction produces a minimum in the
density profile at intermediate distances from the cavity surface on one side
and the condensation of a {\em water drop} in the middle of the pore.
This effect, although quite reasonable, may depend on the model 
and on the simulation technique and, as already stated, deserves a deeper
investigation. 

The observed phenomenon is accompained by the occurence of
a dishomogeneous distribution of the H-bonds throughout the pore.
While the water drop in the middle of the pore seems to have almost the
same distribution of H-bonds as its bulk liquid, the average number of
bonds decreases going towards the interface and its distribution becomes
broader. It is noticeable that even in the shell closest to the interface,
when the density reaches its maximum value, the average number of H-bonds
is much lower than everywhere else. This suggests that the main effect of
the confinement is to disturb the orientational arrangement of molecules
to such an extent that a continuous network of bonds is not favoured up
to distances from the interface of the order of $8 \AA$ approximately.

As far as the intermolecular structure of water is concerned, the
site-site distribution functions give only an average picture of the system.
For this reason the H-bond peak of the $g_{OwHw}(r)$ function is not
dramatically different from that of bulk TIP4P water, nevertheless it is 
clearly apparent from the comparison of all 
the three site-site distribution functions
in Fig.~2 with their counterpart for the bulk liquid that the H-bond network
is strongly distorted and that a number of interstitial (i.e. non H-bonded)
neighbouring molecules must be present. This result is in qualitative 
agreement with the experimental findings of Ref.~\onlinecite{II}, 
although the
distortion of the H-H correlation in that case seems more dramatic.
However we must bear in mind that on one side MD results are model
dependent and on the other side the experimental results were
obtained assuming a weak hydration dependence of the partial structure
factors, an hypothesis which deserves confirmation. As a matter of fact
this MD study will continue to understand
how the water structure depends on the hydration level.      

A further comment on the potential model is mandatory. This model for the
water-substrate interaction seems to be weakly hydrophilic. As a matter of
fact the density in the water layer closest to the substrate increases 
only by the $10 \% $. This may be due to the balance between the
LJ repulsive interaction and the coulombic forces, as suggested also 
by the low number of bonds between water oxygens and protons bonded to nbO.


\newpage

\begin{flushleft}
\begin{table}
\caption{Interaction potential parameters and fractional charges for water
(TIP4P model) and 
silica sites; the locations of water sites 
in the molecular frame are also given  
\label{Table_1}}

\begin{tabular}{lldddddd}
   &    & $\sigma$  &$\epsilon/k_B$ &q     &x        &y        &  z     \\
   &Site&   $(\AA)$ & (K)       &$|e|$& $(\AA)$ & $(\AA)$ & $(\AA)$ \\
\hline
water\tablenotemark[1]&O & 3.154 & 78.0& 0.0 & 0.0 &-0.06556& 0.0 \\
                      &H & 0.0   & 0.0 & 0.52 & 0.75695&0.52032&0.0 \\        
                       &H & 0.0   & 0.0 & 0.52 &-0.75695&0.52032&0.0 \\        
                      &X & 0.0 &0.0 &-1.04& 0.0& 0.0844& 0.0 \\           
silica\tablenotemark[2]& Si&0.0 & 0.0& 1.283& & &  \\
                       & bO&2.70& 230.0& -0.629& & & \\ 
                       & nbO&3.00& 230.0& -0.533& & & \\ 
                       & Hs &0.0 & 0.0& 0.206& & & \\

\end{tabular}
\tablenotetext[1]{TIP4P model~\cite{tip4p}. Label X here
stands for the additional charge site of TIP4P model.}
\tablenotetext[2]{Values from Ref.\ \onlinecite{brodka}; bO: bridging oxygens;
nbO: non bridging oxygens; Hs: protons on the substrate surface.}
\end{table}
\end{flushleft}

\newpage

\noindent
{\bf Figure captions}
\bigskip
\bigskip

\noindent
\textbf{Figure 1} - Computer simulated site-site distribution functions,
calculated according to Eq.~1. Solid line represents distribution
functions obtained by considering water confined in a
interacting silica cavity; dashed line represents distribution
functions of water confined in the same cavity but without any
interaction with the substrate atoms. {\bf a)}~Oxygen-oxygen
distribution function. {\bf b)}~Oxygen-hydrogen distribution
function. {\bf c)}~Hydrogen-hydrogen distribution function.
\bigskip

\noindent
\textbf{Figure 2} - Computer simulated distribution functions,
calculated according to Eq. (2), for confined water (solid lines).
These functions have been corrected taking into account excluded
volume effects, as explained in the text, and are compared with results
obtained for bulk TIP4P water (dashed lines). Due to the
approximations involved in the derivation (see text) the
corrected functions are not
reliable below the minimum approach distances, indicated by the arrow.
{\bf a)}~Oxygen-oxygen distribution function. {\bf b)}~Oxygen-hydrogen
distribution function. {\bf c)}~Hydrogen-hydrogen distribution function.
\bigskip

\noindent
\textbf{Figure 3a} - Computer simulated distribution function of the
subtrate hydrogens ($H_{S}$) with respect to water oxygens ($O_{W}$).
\bigskip

\noindent
\textbf{Figure 3b} - Computer simulated distribution function of the
subtrate oxygens ($O_{S}$) with respect to water hydrogens ($H_{W}$)
(dashed line), and to water oxygens ($O_{W}$) (solid line).
\bigskip

\noindent
\textbf{Figure 4} - Layer analysis of the density profile of confined water
as a function of distance from the center of the pore, calculated with the
substrate-water interaction turned on (solid line) and off (dashed line).
\bigskip

\noindent
\textbf{Figure 5} - Layer analysis of the number of hydrogen bonds ($n_{HB}$)
as a function of distance from the center of the pore, calculated with the
substrate-water interaction turned on (solid line) and off (dashed line). The
orizontal dotted line represents the density-weighted average value of
$n_{HB}$ for confined water with the substrate-water interaction turned
on.
\bigskip

\noindent
\textbf{Figure 6a} - Distributions of hydrogen bonds per molecule
in layers 1, 5, and 9.
\bigskip

\noindent
\textbf{Figure 6b} - Distributions of $cos(\gamma)$, $\gamma$ being the
$H-O\cdots O$ angle
between two H-bonded molecules, in layers 1, 5, and 9.

\newpage
 
\epsfxsize=\hsize \epsfbox{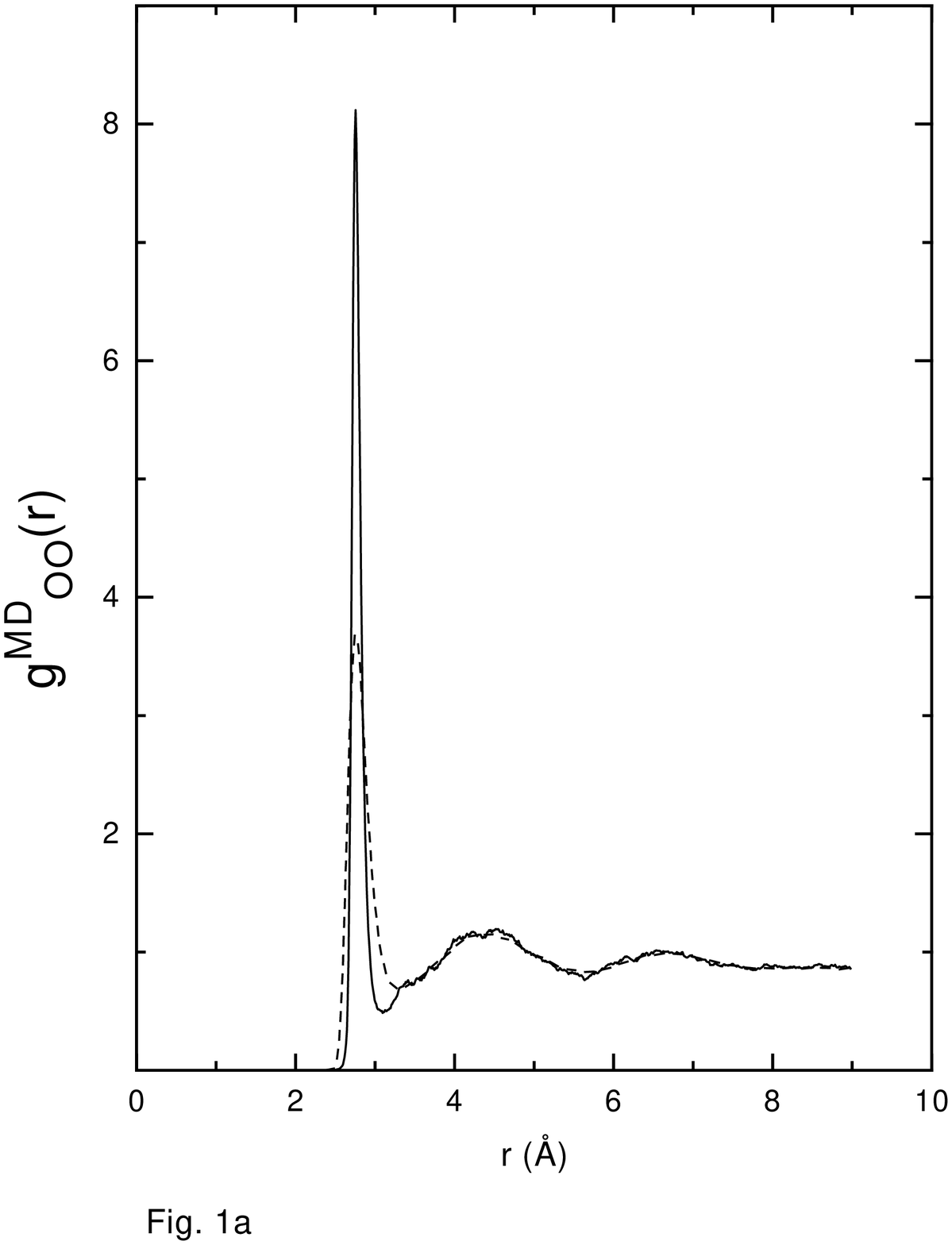}
\epsfxsize=\hsize \epsfbox{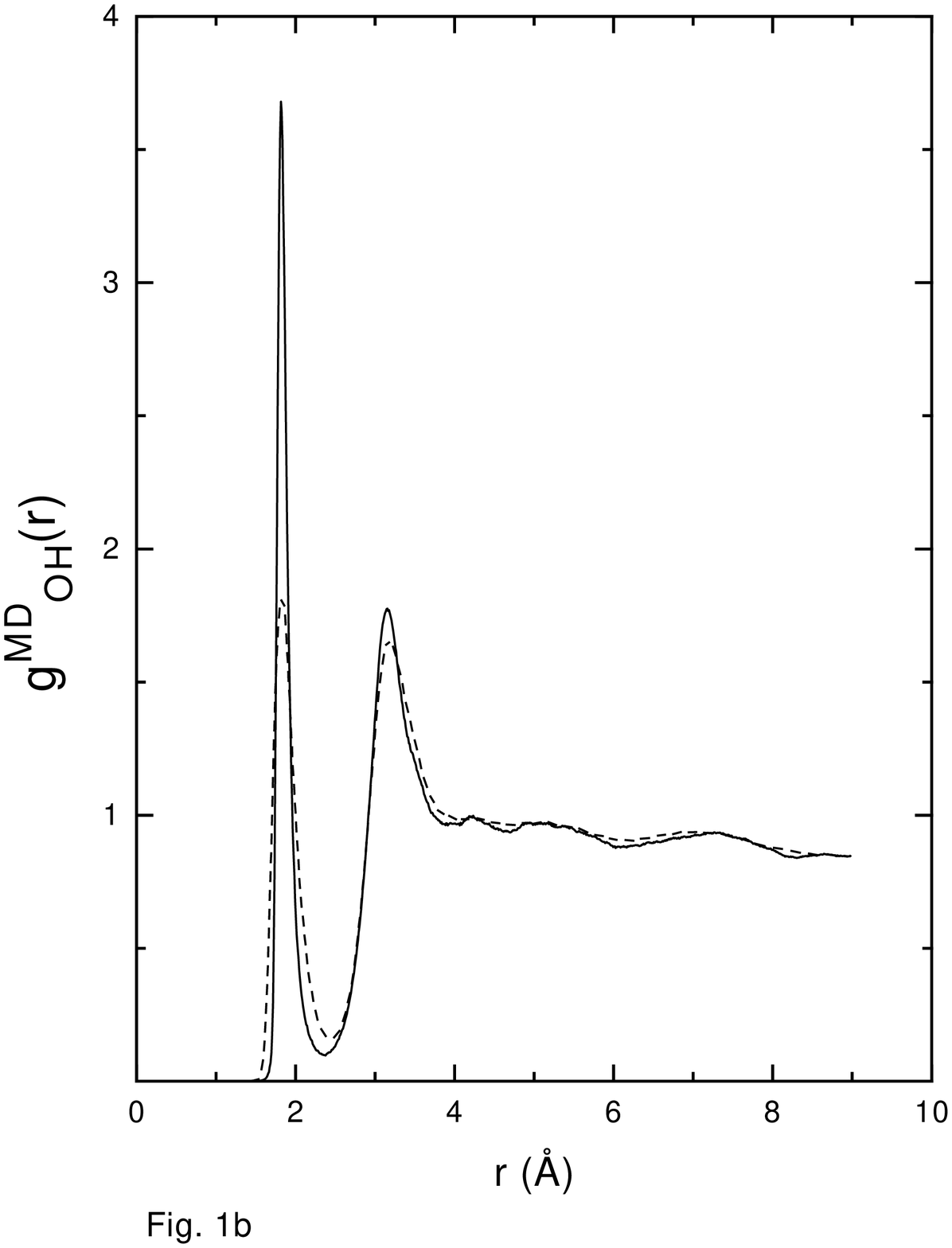}
\epsfxsize=\hsize \epsfbox{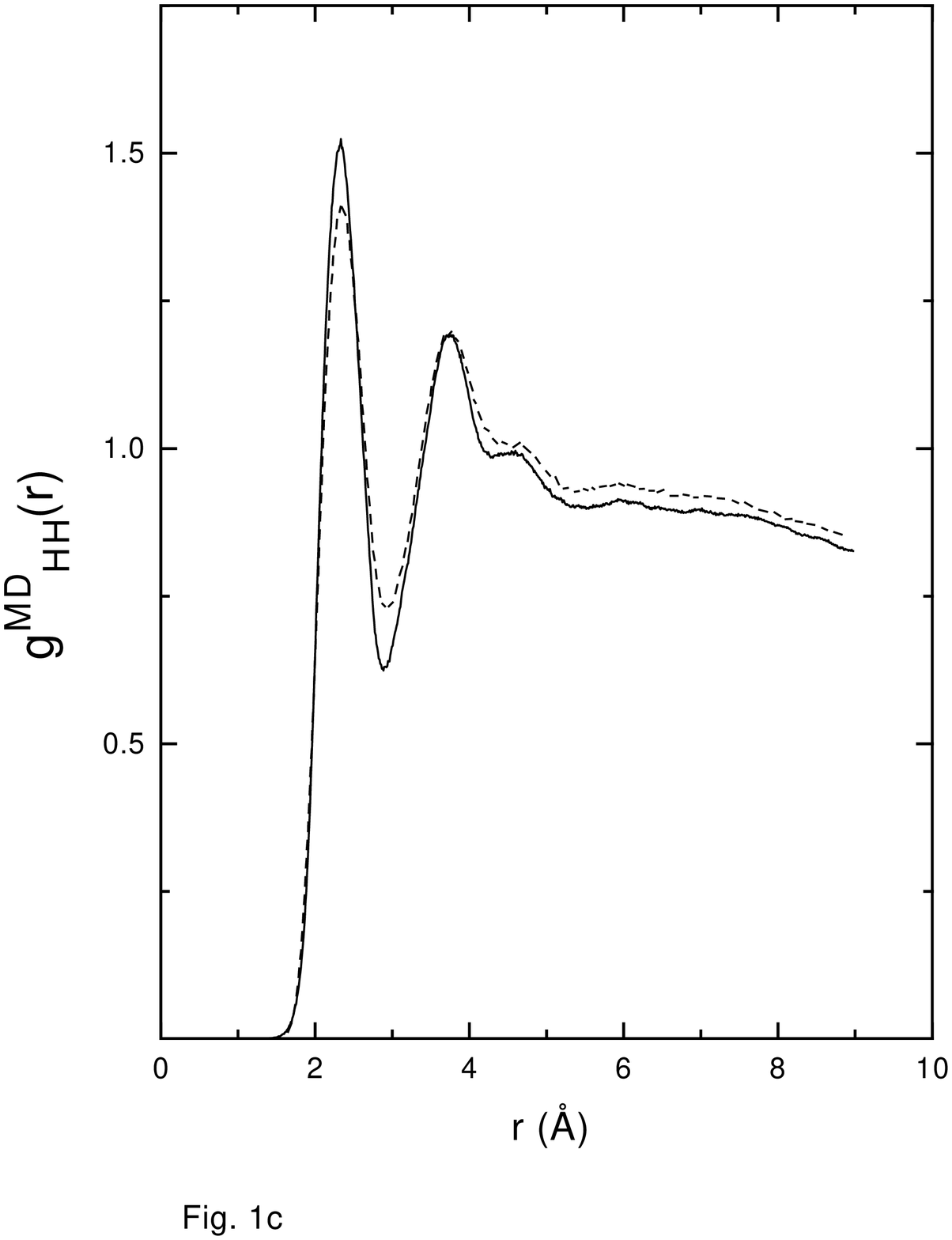}
\epsfxsize=\hsize \epsfbox{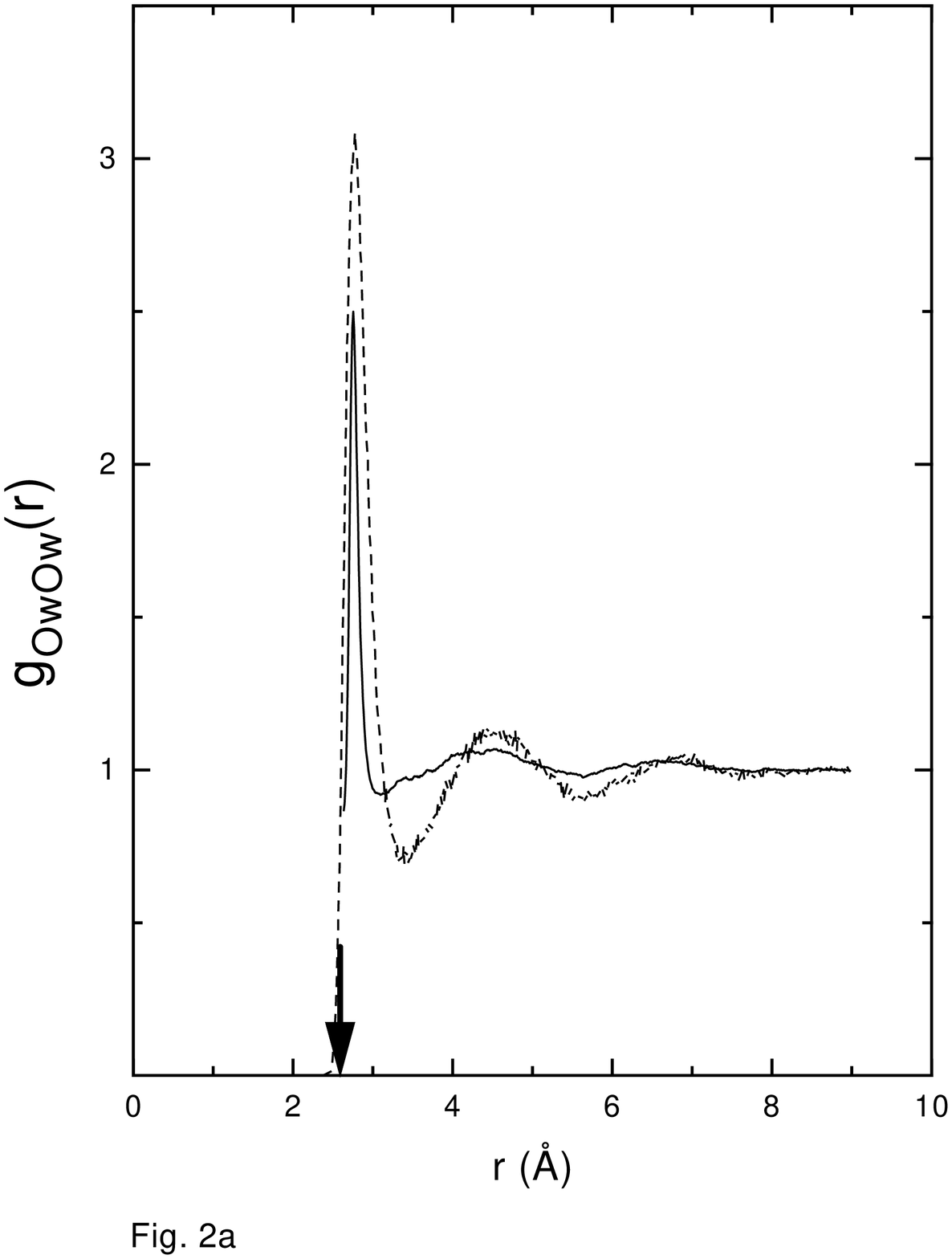}
\epsfxsize=\hsize \epsfbox{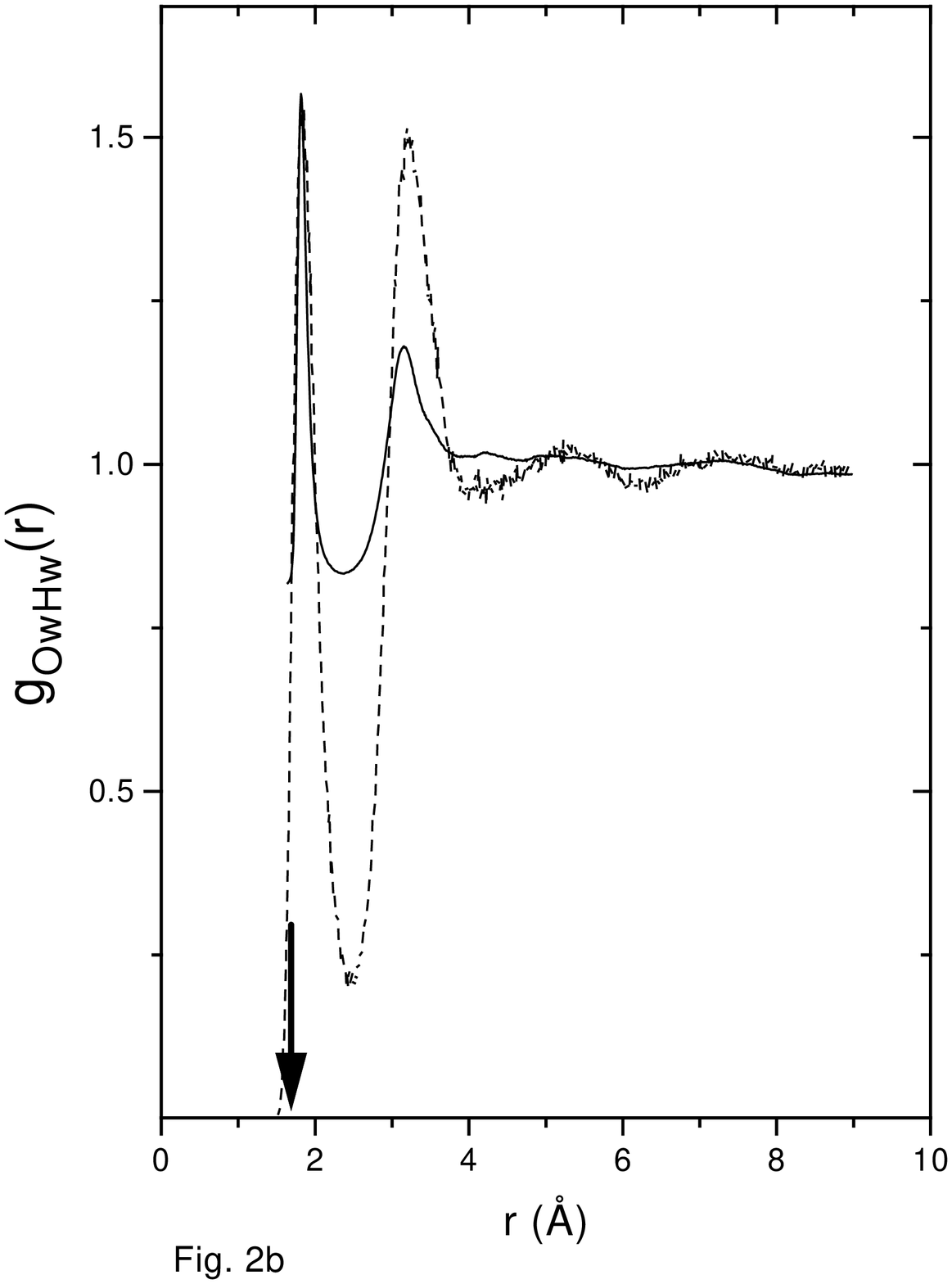}
\epsfxsize=\hsize \epsfbox{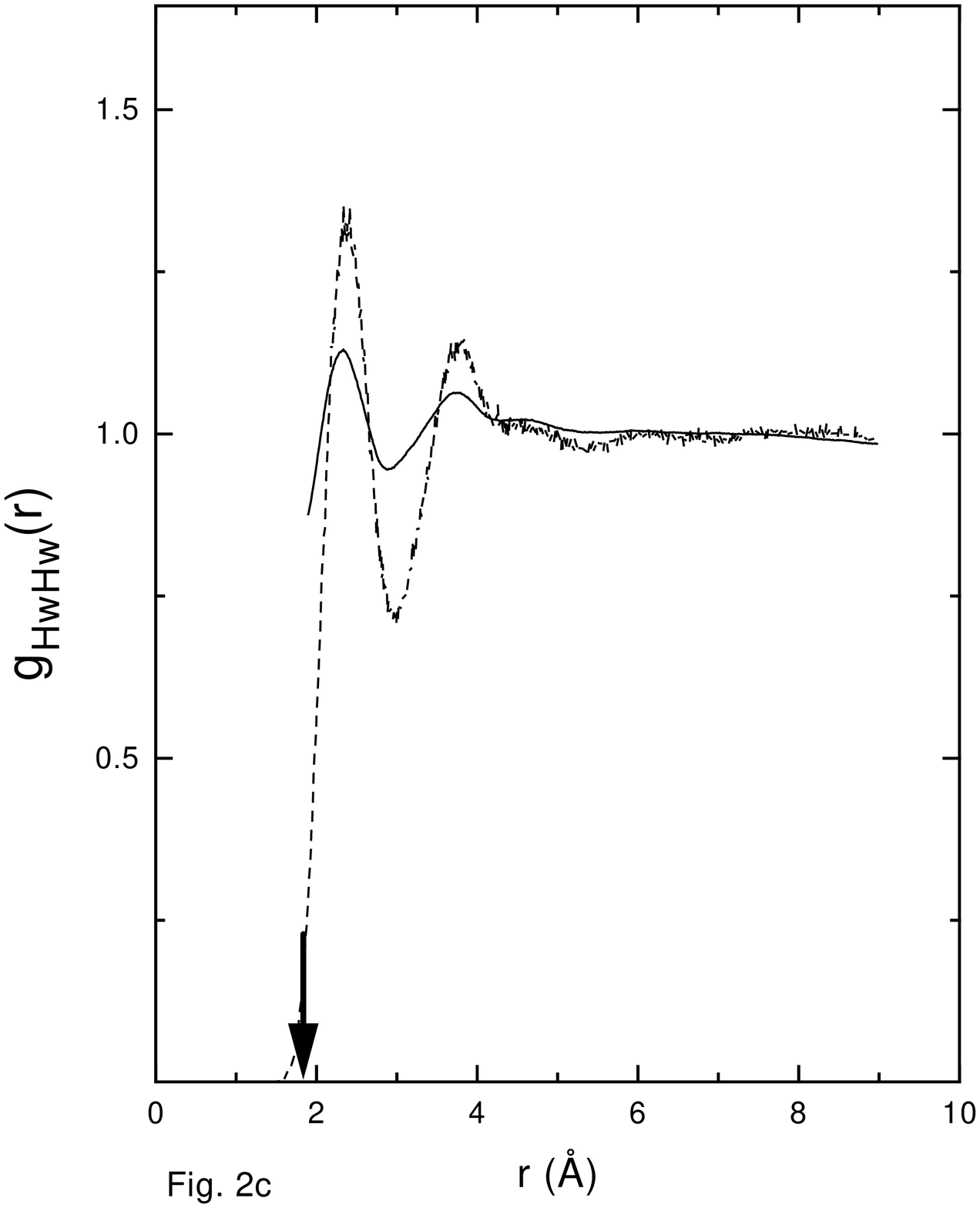}
\epsfxsize=\hsize \epsfbox{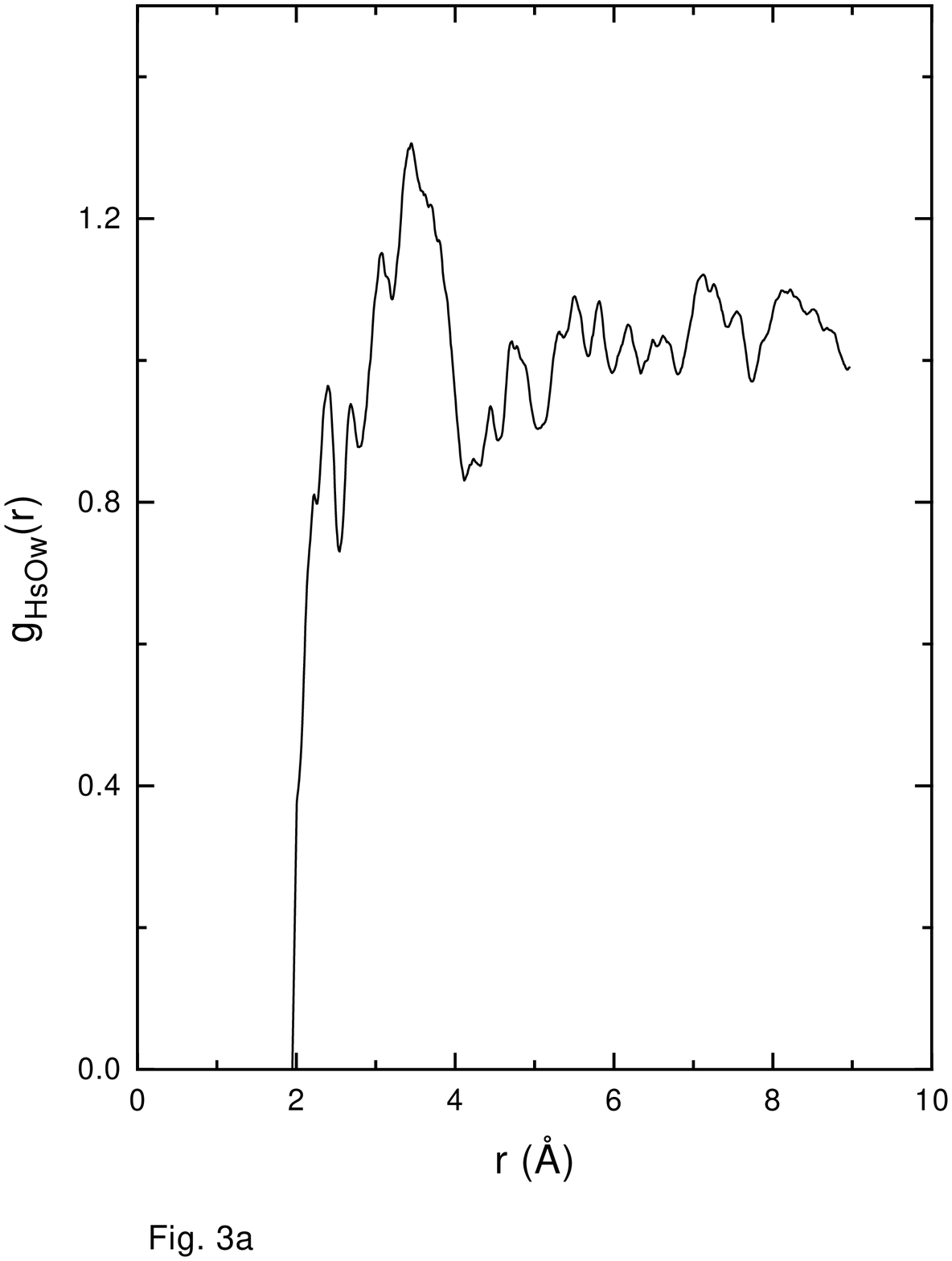}
\epsfxsize=\hsize \epsfbox{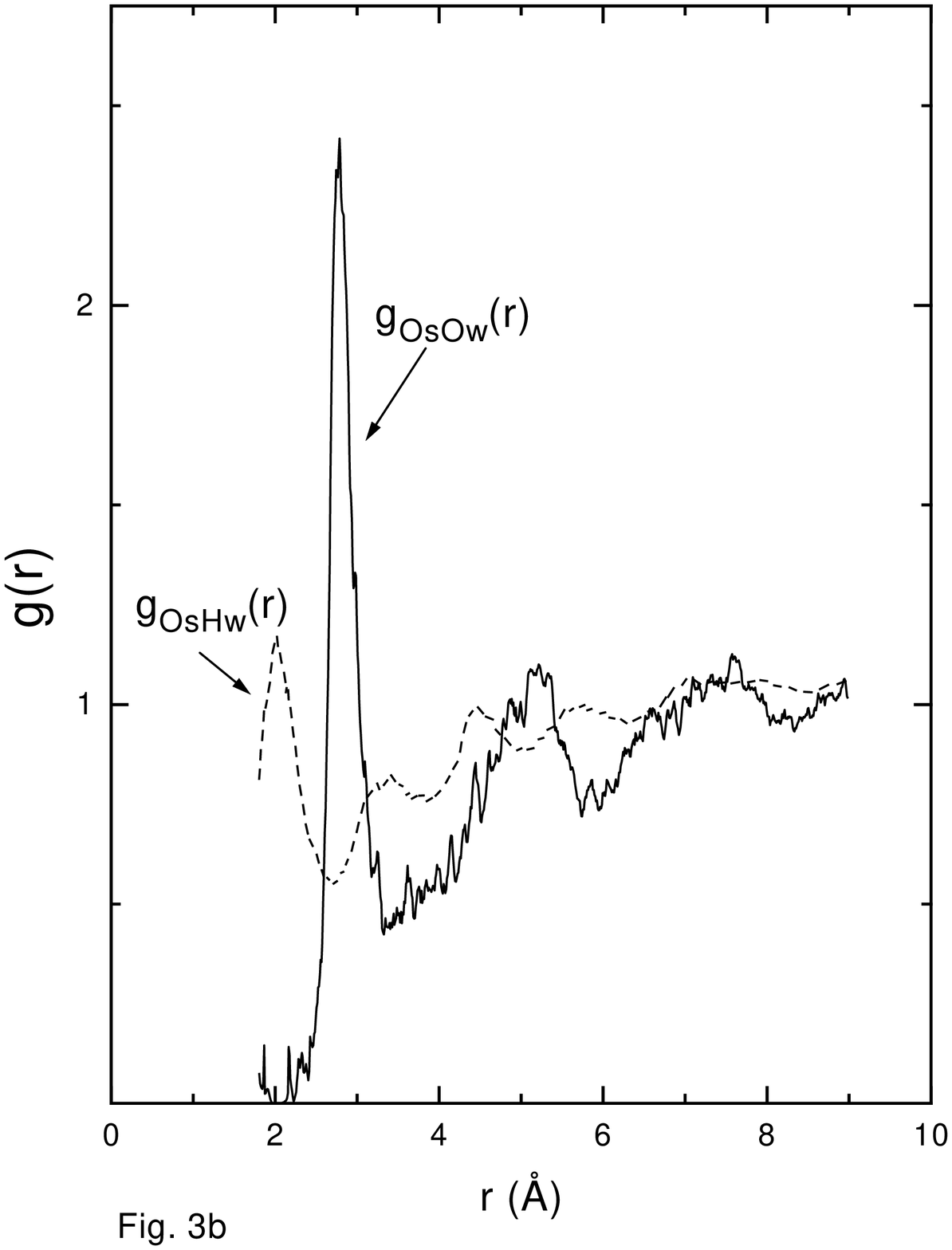}
\epsfxsize=\hsize \epsfbox{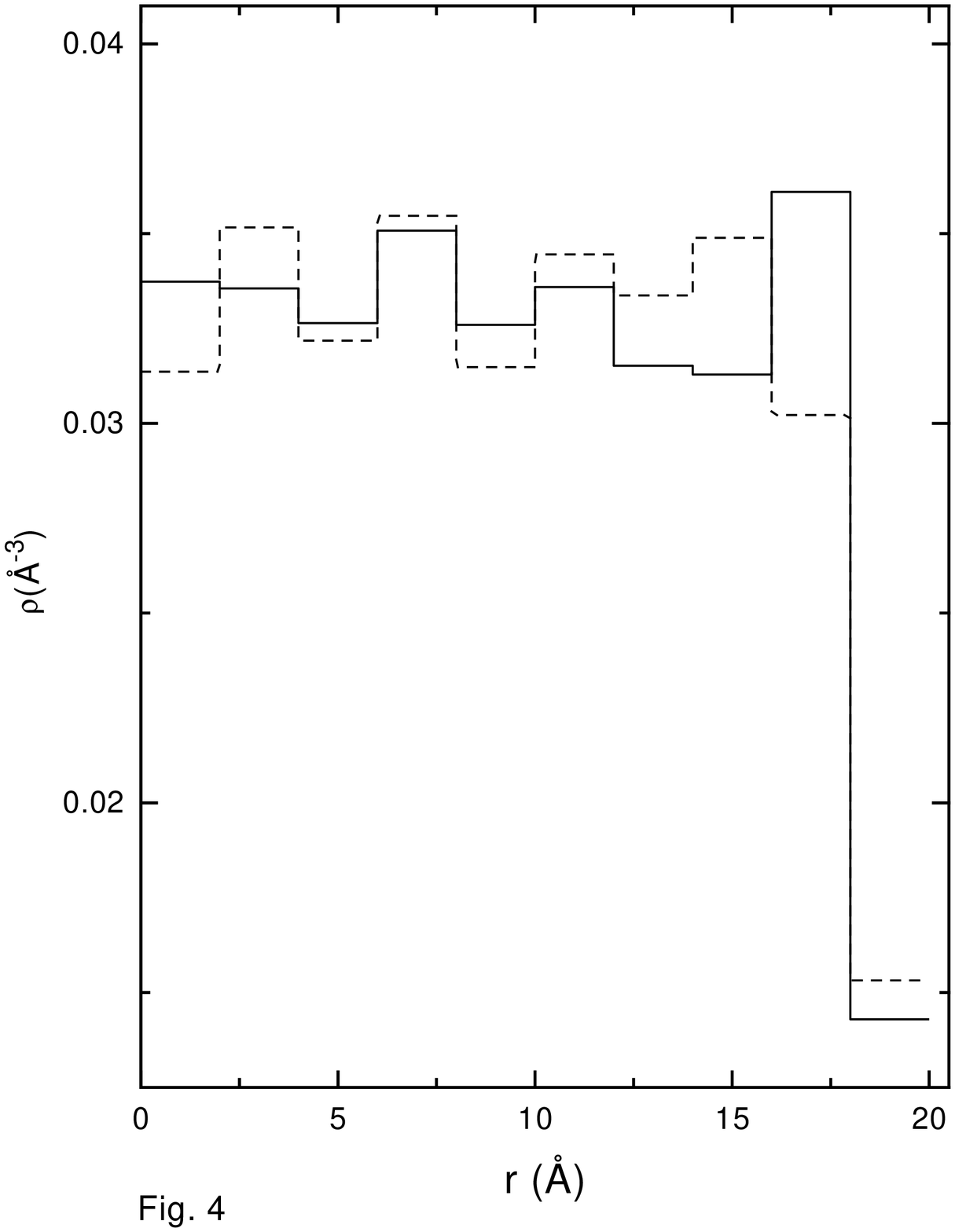}
\epsfxsize=\hsize \epsfbox{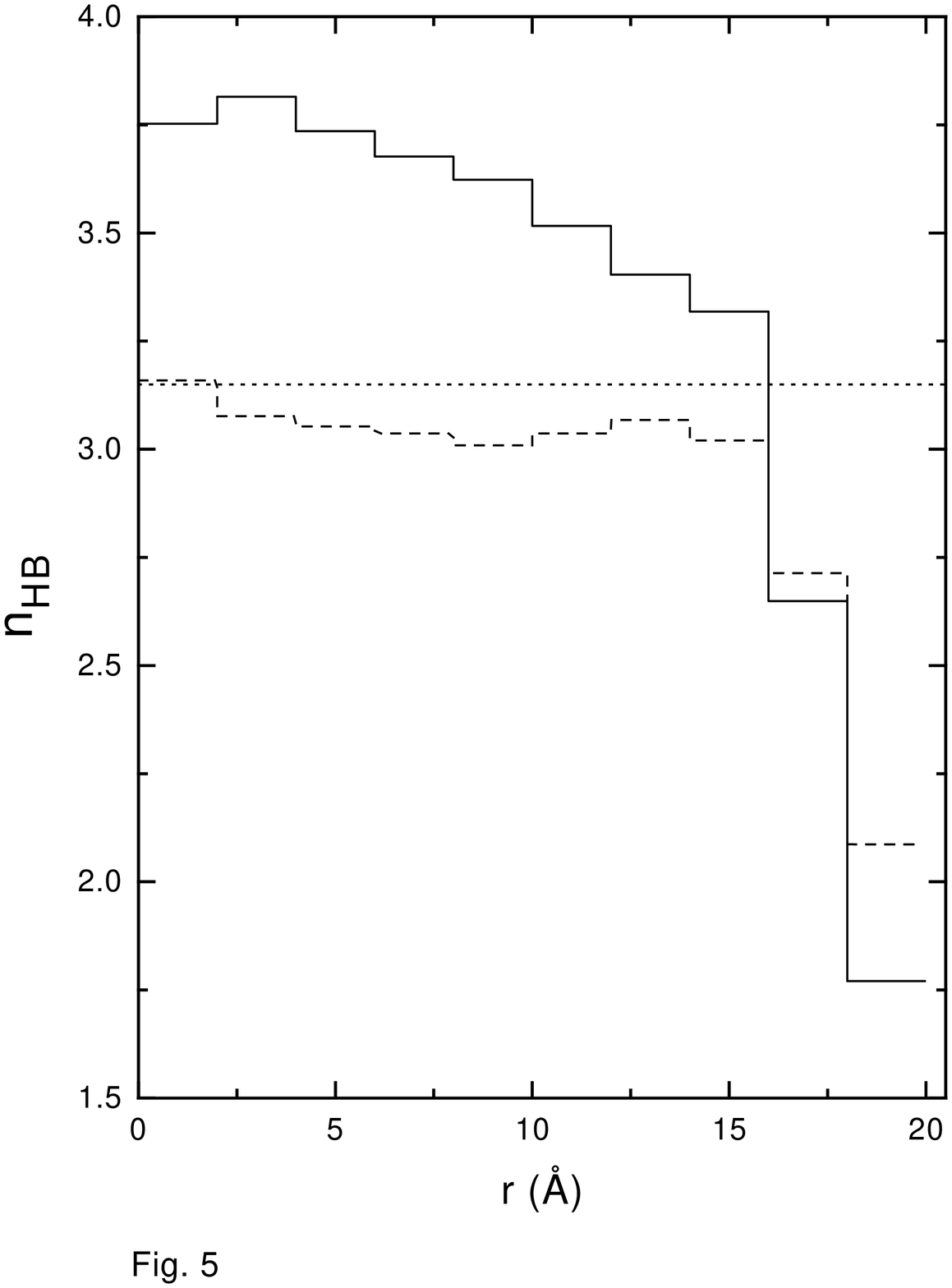}
\epsfxsize=\hsize \epsfbox{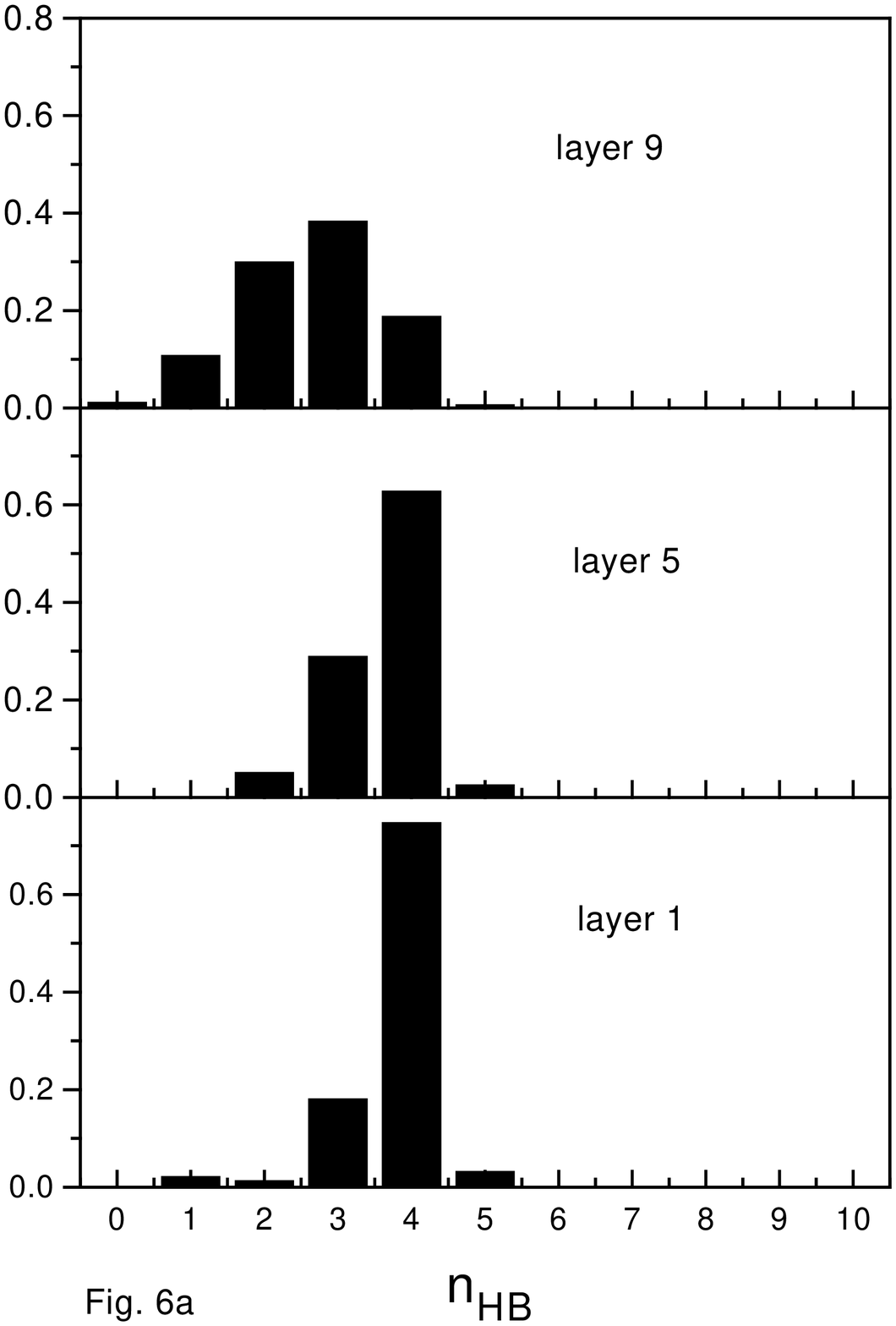}
\epsfxsize=\hsize \epsfbox{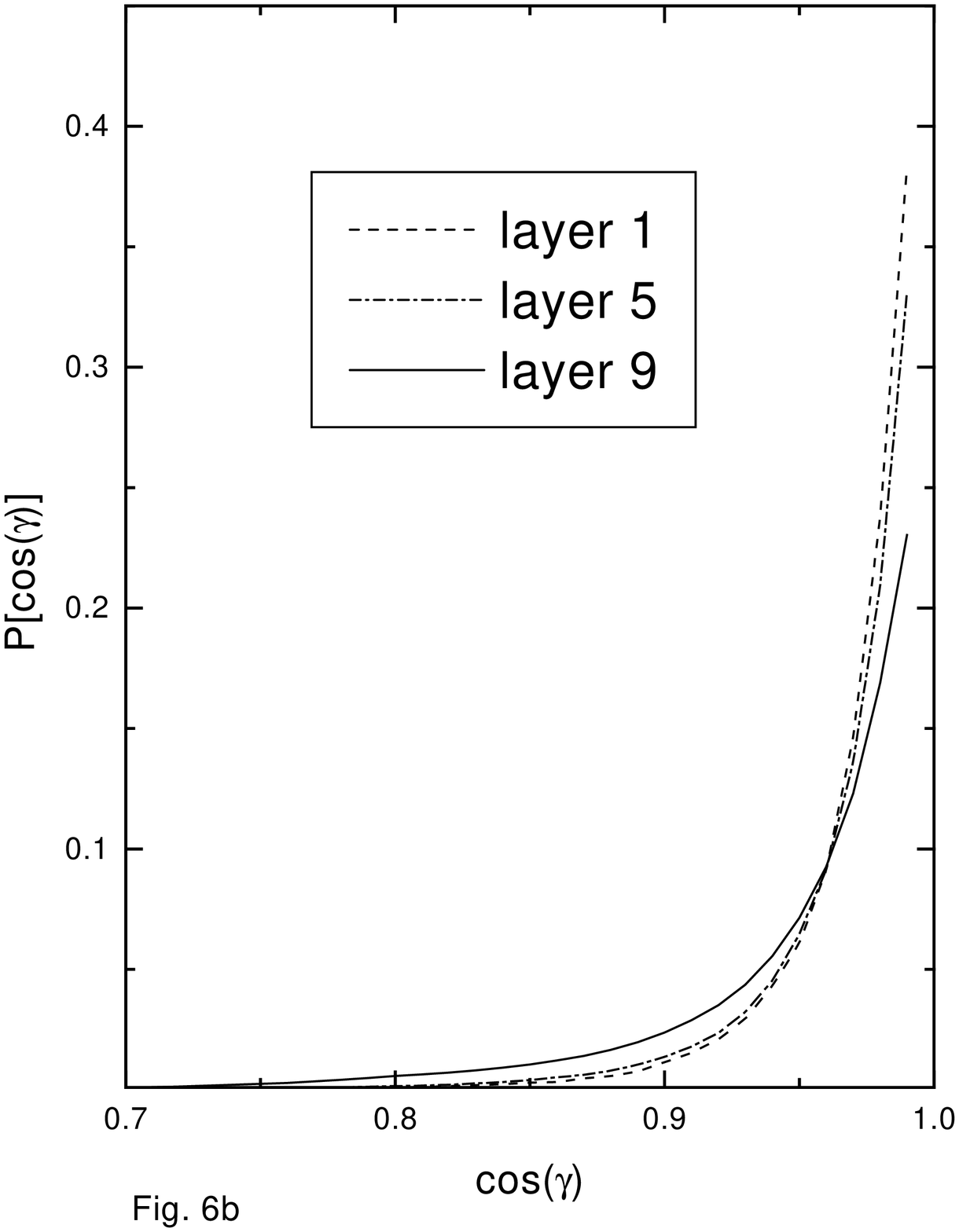}

\end{document}